\documentclass{article}
\usepackage{chngcntr}
\usepackage[utf8]{inputenc}
\usepackage{lmodern}
\usepackage{setspace}
\usepackage{booktabs}       
\usepackage{hyperref}
\usepackage{todonotes}
\usepackage{setspace}
\usepackage{amssymb}        
\usepackage{inputenc}
\usepackage{float}
\usepackage[T1]{fontenc}
\usepackage{chngcntr}
\usepackage{graphicx}       
\usepackage{caption}        
\usepackage{subcaption}     
\usepackage{lineno}         
\usepackage[round]{natbib}         
\usepackage{amsmath}
\usepackage{amssymb}
\usepackage{wrapfig}        
\usepackage[a4paper,left=3cm,right=2cm,top=2.5cm,bottom=2.5cm]{geometry} 
\usepackage[acronym, section=section]{glossaries}
\usepackage{textgreek}
\setacronymstyle{long-short}
\makeindex
\graphicspath{ {Figures/} }
\linenumbers
\graphicspath{ {Figures/} }
\usepackage[labelfont=bf]{caption}
\usepackage[labelsep=period]{caption}
\usepackage{xr}

\begin{document}
\nolinenumbers
\begin{titlepage}
   \begin{center}
       \vspace*{0.2cm}
       \centering
       \LARGE
       \hspace*{1.3cm}
       \textbf{Detecting post-stroke aphasia} 
       \newline
       \hspace*{0.1cm}
      \textbf{ using EEG-based neural envelope tracking} 
      \newline
      \hspace*{2.5cm}
      \textbf{of natural speech}
       \newline

       \vspace{1cm}
       \large
       Pieter De Clercq\textsuperscript{1}, Jill Kries\textsuperscript{1},  Ramtin Mehraram\textsuperscript{1}, Jonas Vanthornhout\textsuperscript{1}, Tom Francart\textsuperscript{1} and Maaike Vandermosten\textsuperscript{1}

       \textsuperscript{1}: Experimental Oto-Rhino-Laryngology, Department of Neurosciences, Leuven Brain Institute, KU Leuven, Belgium\\
       \vspace{2cm}
        
        \textbf{Version:} \\ March 2023 \\
       \vspace{2cm}
        
        \textbf{Corresponding author:}\\Pieter De Clercq (pieter.declercq@kuleuven.be)
    \end{center}

    \vfill
    \newpage
\end{titlepage}
\begin{center}
 \section*{Abstract}
\end{center}
\noindent
After a stroke, approximately one-third of patients suffer from aphasia, a language disorder that impairs communication ability. The standard behavioral tests used to diagnose aphasia are time-consuming, require subjective interpretation, and have low ecological validity. As a consequence, comorbid cognitive problems present in individuals with aphasia (IWA) can bias test results, generating a discrepancy between test outcomes and everyday-life language abilities. Neural tracking of the speech envelope is a promising tool for investigating brain responses to natural speech. The envelope of speech is crucial for speech understanding, encompassing cues for detecting and segmenting linguistic units, e.g., phrases, words and phonemes. In this study, we aimed to test the potential of the neural envelope tracking technique for detecting language impairments in IWA.

\noindent
We recorded EEG from 27 IWA in the chronic phase after stroke and 22 healthy controls while they listened to a 25-minute story. We quantified neural envelope tracking in a broadband frequency range as well as in the delta, theta, alpha, beta, and gamma frequency bands using mutual information analysis. Besides group differences in neural tracking measures, we also tested its suitability for detecting aphasia at the individual level using a Support Vector Machine (SVM) classifier. We further investigated the required recording length for the SVM to detect aphasia and to obtain reliable outcomes.

\noindent
IWA displayed decreased neural envelope tracking compared to healthy controls in the broad, delta, theta, and gamma band, which is in line with the assumed role of these bands in auditory and linguistic processing of speech. Neural tracking in these frequency bands effectively captured aphasia at the individual level, with an SVM accuracy of 84\% and an area under the curve of 88\%. Moreover, we demonstrated that high-accuracy detection of aphasia can be achieved in a time-efficient (5 minutes) and highly reliable manner (split-half reliability correlations between R=0.62 and R=0.96 across frequency bands).

\noindent
Our study shows that neural envelope tracking of natural speech is an effective biomarker for language impairments in post-stroke aphasia. We demonstrated its potential as a diagnostic tool with high reliability, individual-level detection of aphasia, and time-efficient assessment. This work represents a significant step towards more automatic, objective, and ecologically valid assessments of language impairments in aphasia.

\vspace{1cm}
\noindent
\textbf{Keywords: }Aphasia, natural speech processing, neural envelope tracking, diagnostics
\newpage
\section{Introduction}\label{Introduction}
Aphasia is an acquired language disorder impairing communication ability and is principally caused by a stroke in the language-dominant left hemisphere \citep{Papathanasiou2017}. The current practice is to diagnose aphasia by means of behavioral language tests. However, these tests suffer from influences of co-morbid motor and cognitive problems \citep{Rohde2018} and of a low ecological validity \citep{Devanga2021, Wallace2013}. Novel analysis techniques for EEG-data, i.e., neural tracking of speech (e.g., see \cite{Lalor2009, Brodbeck2022, Crosse2021}), allow measuring brain responses while participants listen to natural speech, providing an ecologically valid way to measure speech processing. In this study, we test the potential of these novel EEG analyses for detecting language impairments in aphasia with high accuracy and in a time-efficient way.

The current standard for diagnosing aphasia is based on performance on behavioral language tests, such as the Western Aphasia Battery \citep{Kertesz1982}, the Token test \citep{deRenzi1978} or a picture-naming test \citep{nbt}. Yet, these tests have several disadvantages. First, behavioral testing is time-consuming, requiring active cooperation of the patient and scoring by the clinician. Second, behavioral assessment can lead to inaccurate initial diagnoses due to concomitant motor, memory, attention and executive impairments \citep{Rohde2018}, reportedly affecting over 80\% of individuals with aphasia (IWA) \citep{ElHachioui2014}. Finally, language tests consist of rather artificial tasks in which sounds, phonemes, words or short sentences are presented in isolation. This contrasts with natural speech processing where language components interact, and higher-level context integration takes place \citep{Hamilton2018, Kandylaki2019}. Consequently, there is a discrepancy between clinical assessment and a patient's natural speech abilities in everyday life \citep{Lesser1995, Kim2022, Stark2021, Wallace2013}.

EEG-based event-related potential (ERP) studies have been conducted to address limitations of behavioral testing. Studies have shown that IWA exhibit altered ERP components such as the P1, N1, P2, N2, P300, and N400 in response to language stimuli \citep{Aerts2015, Becker2007, Ilvonen2001, Ofek2013, Pulvermuller2004, Robson2017}. Together with the potential for automatic assessment that requires less active participation from the patient, these alterations suggest that ERPs may have diagnostic value in aphasia \citep{Cocquyt2020}. Nonetheless, ERP paradigms involve artificial language stimuli presented repeatedly to the participant, which questions the ecological validity of the obtained outcomes \citep{Le2018}. Furthermore, previous ERP studies in aphasia have not reported on the reliability of ERPs, the minimal required recording length and the sensitivity to capture language impairments at the individual level.

Recent studies have investigated the EEG response to natural, running speech, which could open new perspectives to studying natural speech processing in IWA. When listening to speech, the brain tracks the temporal envelope, which contains essential cues for speech understanding. The envelope consists of the slow-varying temporal modulations in the speech signal and encompasses cues for detecting and identifying lexical units (i.e., phonemes, syllables, words and phrases) and prosody \citep{Peelle2012}. In fact, previous research showed that listeners can understand speech based on the low-frequency temporal envelope only \citep{Shannon1995}. Neural envelope tracking can be measured by applying encoding and decoding models on the stimulus and the recorded EEG. The level of tracking is reflected in the extent to which the models can either predict the neural signals or decode the envelope. In a linear \citep{Crosse2021} or a mutual information-based \citep{DeClercq2023} model, the neural tracking outcomes can be visualized over time and space (i.e., EEG channels), obtaining response properties similar to traditional ERP components \citep{Brodbeck2022}. The neural tracking technique is rapidly evolving and has led to crucial new insights as to how natural speech is processed in the brain. Neural envelope tracking is strongly related to speech understanding \citep{Ding2013, Etard2019, Kaufeld2020}, and can be used to objectively quantify speech intelligibility \citep{Vanthornhout2018, Gillis2022}.

Prior research assigned the low-frequency temporal envelope primarily to speech understanding. The low-frequency envelope, i.e., delta (0.5–4~Hz) and theta (4-8~Hz) band, encompass cues for detecting and segmenting lexical units. The theta band tracks syllables and lower-level acoustic processing of speech \citep{Etard2019}, while the delta band signal is associated with processing speech prosody and segmenting higher-level linguistic structures such as words and phrases \citep{Ding2016,Giraud2012,Kaufeld2020}. In addition to the delta and theta band, reflecting synthesis of higher-level auditory and linguistic structures,  the alpha and beta bands are involved in attention and auditory-motor coupling \citep{Wostmann2017, Fujioka2015}, while the gamma band is involved in encoding phonetic features \citep{Giraud2012, Gross2013, Hyafil2015}. In conclusion, specific frequency bands are believed to reflect different stages of speech processing.

Neural envelope tracking of natural speech has been investigated in several clinical populations with language impairments. For individuals with primary progressive aphasia, a language disorder caused by a neurodegenerative disease, \citet{Dial2021} reported increased neural tracking in the theta band but no group differences in the delta band. The researchers argued that enhanced theta band tracking in individuals with primary progressive aphasia might reflect a compensation mechanism through increased reliance on acoustic cues. For individuals with dyslexia, a disorder characterized by phonological processing difficulties, decreased tracking in delta, theta and beta/gamma (phoneme- and phonetic-level) band have been reported \citep{DiLiberto2017,Lizarazu2021,Mandke2022}. In conclusion, these studies have shown the potential for neural tracking to capture language impairments in clinical cohorts.

The present study investigated whether we can differentiate IWA in the chronic phase after stroke (i.e., $\geq$ 6 months post-stroke) from neurologically healthy, age-matched controls using EEG-based neural envelope tracking. Specifically, we used mutual information analyses to quantify neural envelope tracking, which captures linear and nonlinear effects and outperforms linear models \citep{DeClercq2023}. We described both groups' responses to the speech envelope temporally and spatially at broadband frequency range. We further investigated neural tracking in specific frequency bands ranging from delta to gamma band, as different frequency bands are involved in different (sub-)lexical processes \citep{Etard2019, Ding2016, Giraud2012, Keitel2018, Peelle2012}. 

Secondly, we assessed the suitability of the neural tracking technique as a biomarker to capture language processing difficulties. To this end, we used a Support Vector Machine (SVM) to classify participants as healthy or aphasic using MI measures in different frequency bands as input to the model. Finally, we investigated how much data the neural tracking technique requires for good classification and reliable outcomes.  

\section{Materials and methods}
\subsection{Participants}
Our sample comprised 27 IWA (seven female participants, 73 $\pm$11 y/o) in the chronic phase ($\geq$ 6 months) after stroke and 22 neurologically healthy controls (seven female participants, 72 $\pm$7 y/o). There was no significant age difference between groups (unpaired Wilcoxon rank sum test: W=343.5, p=0.36). IWA were recruited at the stroke unit of the University Hospital Leuven and via speech-language pathologists. Healthy controls were recruited, making sure they matched the age of IWA at the group level. The inclusion criteria for IWA were: (1) a left-hemispheric or bilateral stroke, (2) a diagnosis of aphasia in the acute stage after stroke using behavioral language tests and (3) no formal diagnosis of a psychiatric or neurodegenerative disorder. A detailed overview including demographic information (age, sex, time since stroke, speech-language therapy, performance on diagnostic tests for aphasia at time of participation) and lesion information (stroke type, affected blood vessel, lesioned hemisphere) about the aphasia sample can be found in the Supplementary materials (Supplementary Table \ref{table:participants}). For more information regarding recruitment strategy and diagnosis in the acute stage after stroke, we refer to \citet{Kries2022}. The study was approved by the ethical committee UZ/KU Leuven (S60007), and all participants gave written consent before participation.  Research was conducted in accordance with the principles embodied in the Declaration of Helsinki and in accordance with local statutory requirements.

Participants completed standardized clinical tests for aphasia at the time of participation as described in detail in \citet{Kries2022}. IWA scored significantly lower on the 'Nederlandse Benoemtest', i.e., Dutch Naming Test \citep{nbt}, and the ScreeLing test \citep{ElHachioui2017, VischBrink} compared to healthy controls (W=57.5, p<0.001; W=101, p<0.001, respectively). Although seven IWA did not score below the cut-off threshold for aphasia on either of these tasks, they were still attending speech-language therapy sessions at the time of participation and had extended documentation of language deficits in the acute stage after stroke \citep{Kries2022}. 

\subsection{EEG experiment}
The EEG measurements took place in a soundproof, electromagnetically shielded booth using a 64-channel BioSemi ActiveTwo system (Amsterdam, the Netherlands) at a sampling frequency of 8,192~Hz. Participants were instructed to listen to a 25-minute-long story, \textit{De Wilde Zwanen}, written by Christian Andersen and narrated by a female Flemish-native speaker, presented in silence while their EEG was recorded. The story was cut into five parts with an average duration of 4.84 minutes. After each story part, participants answered content questions about the preceding part, introduced to make the participant follow the content attentively. As one participant fell asleep during two story parts, we decided to leave these parts out of the analysis for that participant (sub-019, Supplementary Table \ref{table:participants}). Participants had a short break after each story part and answered content questions about the preceding part. The protocol introduced these questions to make participants follow the story attentively. The story was presented bilaterally through ER-3A insert earphones (Etymotic Research Inc, IL, USA) using the software platform APEX \citep{Francart2008}. 

We determined a subject-dependent intensity level at which the story was presented based on the thresholds of the pitch tone audiometry (PTA). We defined hearing thresholds for octave frequencies between .25 and 4~kHz. For normal hearing participants, the story was presented at 60 dBA. For hearing impaired participants, defined as participants that have a hearing threshold >25 $\>$dB hearing loss on frequencies below 4~kHz, the volume was augmented with half of the pure tone average of the individual thresholds at .25, .5 and 1~kHz for both ears individually. This procedure was adapted from \citet{Jansen2012}. To check whether age-related hearing loss differed between both groups, we calculated the Fletcher index, i.e., average of PTA thresholds at .5, 1 and 2 kHz. Hearing levels did not differ between groups (Fletcher index averaged across the right and left ear: W=326.5, p=0.56).

\subsection{Signal processing}
\subsubsection*{Envelope extraction}
We used a gammatone filter bank \citep{Sondergaard2012} to extract the envelope. We used 28 channels spaced by one equivalent rectangular bandwidth and center frequencies from 50~Hz until 5000~Hz. The envelopes were extracted from each sub-band by taking the absolute value of each sample and raising it to the power of 0.6. The resulting 28 sub-band envelopes were averaged to obtain a single envelope. Next, the envelope was downsampled to 512~Hz to decrease processing time. The envelope was then filtered in frequency ranges of interest. These include delta (0.5-4~Hz), theta (4-8~Hz), alpha (8-12~Hz), beta (12-30~Hz), low-gamma (30-49~Hz) and a broad (0.5-49~Hz, including all individual frequency ranges) band. We used high- and lowpass filters, with a transition band of 10$\%$ below the highpass and 10$\%$ above the lowpass frequency. A Least Squares filter of order 2000 was used, and we compensated for the group delay. After filtering, the envelope was normalized and further downsampled to 128~Hz.

\subsubsection*{EEG data processing}
EEG data were pre-processed using the Automagic toolbox \citep{Pedroni2019} and custom Matlab scripts (The MathWorks Inc., Natick, MA, USA, 2021). The EEG signals were first downsampled to 512~Hz to decrease processing time. Artifacts were removed using the artifact subspace reconstruction method \citep{Mullen2015}. Next, an independent component analysis was applied to the data, and components classified as "brain" or "other" (i.e., mixed components), using the EEGLAB plugin ICLabel \citep{Pion2019}, with a probability higher than 50$\%$ were preserved (average number of removed components: 26 $\pm$7). The neural signals were projected back to the channel space, where the signals were average referenced. Subsequently, we filtered the EEG data in the same frequency bands using the same Least Squares filter as in the envelope extraction method. Next, normalization and further downsampling to 128~Hz were applied.

\subsection{Neural envelope tracking}
We investigated neural envelope tracking using the Gaussian copula MI analysis \citep{Ince2017}. In the Gaussian copula approach, all variables (the envelope and EEG channels) are first ranked on a scale from 0 to 1, obtaining the cumulative density function (CDF). By computing the inverse standard normal CDF, the data distributions of all variables are transformed to perfect standard Gaussians. Subsequently, the parametric Gaussian MI estimate can be applied to the data provided by:

\begin{equation}\label{eq:1}
I(X;Y) = \frac{1}{2ln2}\Biggl[\frac{|\sum_{X}||\sum_{Y}|}{|\sum_{XY}|}\Biggr]
\end{equation}
where I(X;Y) equals the MI between X and Y (here, the EEG and the envelope), expressed in bits. $|\sum_{X}|$ and $|\sum_{Y}|$ are the determinants of the covariance matrices of $X$ and $Y$, and $|\sum_{XY}|$ is the determinant of the covariance matrix for the joint variable. To obtain temporal information on MI, we shifted the EEG as a function of the envelope over time (using an integration window -200 to 500 ms) and applied Eq. (\ref{eq:1}) at each sample. The result forms the temporal mutual information function (TMIF) and reflects how the brain processes speech over time \citep{DeClercq2023, Zan2020}. For an in-depth explanation of the Gaussian copula MI method, we refer to \citet{Ince2017}. For a more practical explanation of the TMIF in the context of neural envelope tracking, see \cite{DeClercq2023}.

We calculated the single-channel TMIF and the multivariate TMIF, analog to a (linear) encoding and decoding model, respectively. The single-channel TMIF calculates the TMIF for each channel individually, providing both temporal (i.e., peak latency and peak magnitude) and spatial (i.e., topography) information on speech processing. Alternatively, the multivariate TMIF determines the multivariate relationship between multiple EEG channels combined and the speech envelope. This latter method is statistically more powerful as it takes interactions between EEG channels into account. However, it is restricted to temporal interpretations only. For the multivariate TMIF, we used a channel selection including fronto-central and parieto-occipital channels that contribute to speech processing \citep{Lesenfants2019}. Our channel selection is visualized in Supplementary Fig. \ref{fig:ch_s}. 

\subsubsection*{Permutation testing}
Neural tracking (MI in bits, in this case) is a relative metric and should be compared to a null-distribution to quantify the meaningfulness of the derived values \citep{DeClercq2023}. We created stationary noise that matched the spectrum of the envelope per frequency band individually. Next, we calculated the MI between the noise envelope and the EEG per participant and repeated this process 1000 times. The significance level was then determined as the 95th percentile of permutations per participant (resulting in a single significance level per participant).

No significant differences were found in the significance level between IWA and controls for any frequency band, as determined by Wilcoxon rank sum tests. Supplementary Fig. \ref{fig:sigs} displays the significance levels of the multivariate TMIF for all frequency bands categorized by group. As no significant differences in the significance level were found between groups, we used a single significance level (i.e., the 95th percentile of permutations across all participants) to interpret the multivariate TMIFs in the Results section.

\subsection{Statistics}
\subsubsection*{Group comparisons}
We compared neural envelope tracking for IWA with the control group for broadband as well as for delta, theta, alpha, beta and gamma frequency ranges. For the single-channel TMIF, we performed non-parametric spatio-temporal cluster-based permutation tests \citep{Maris2007}, indicating clusters in the TMIF over time and space with the largest group difference at threshold p<0.05. For the multivariate TMIF, we performed non-parametric temporal cluster-based permutation tests \citep{Maris2007}, indicating clusters of samples with the largest group difference at threshold p<0.05. 

\subsubsection*{Support Vector Machine Classification}
We investigated whether EEG-based envelope tracking outcomes can be used for detecting aphasia. To this end, we used a Support Vector Machine (SVM) to classify held-out participants as control or aphasic using the Scikit-Learn (v. 0.24.2) library in Python \citep{Pedregosa2011}. The multivariate TMIFs for all five individual frequency bands (delta to gamma) were used as input to the model. Additionally, we added age of the participant, as it influences neural envelope tracking \citep{Decruy2019}. We chose a radial basis function kernel SVM and performed a nested cross-validation approach. In the inner cross-validation, the C-hyperparameter and pruning (i.e., length of the TMIFs) were optimized (accuracy-based) and tested in a validation set using 5-fold cross-validation. The trained model was then tested on the test set, for which we used a leave-one-subject-out cross-validation approach.

The performance of the SVM classifier was evaluated by computing the receiver operating characteristic (ROC) curve and calculating the area under the curve (AUC). We further reported the overall accuracy, the F1-score, the sensitivity and the specificity of the classifier.

\textbf{Feature contribution. }To obtain a proxy for the relevant contribution of each frequency band, we left out a single band and re-fitted the SVM. We repeated this process for all five frequency bands and reported the corresponding performance drop (AUC, accuracy, F1-score).

\subsubsection*{Recording time}
\textbf{Classification. }From a practical perspective, we were interested in how much data the neural envelope tracking technique requires to detect aphasia accurately and obtain stable, reliable results. We iteratively cropped the EEG recording and the envelope in steps of 2 minutes (using the first 1 minute, first 3 minutes, 5, 7… up to the entire 25 minutes of recording time) and calculated the TMIF per frequency band per time duration. Next, we investigated the amount of minutes required for the SVM to reach its classification potential. As described above, we trained and tested our SVM per time duration in the same fashion as the entire duration. Performance (AUC, accuracy, F1-score) was plotted as a function of recording time. We determined the knee point, i.e., the point at which the performance benefit starts to saturate, using the "kneed" python package \citep{Satopaa2011}. The knee point of this curve reflects the point at which the increase in model performance may no longer be worth the corresponding effort.

\textbf{Within-subjects stability. }Second, we investigated the data required to obtain stable, reliable results. We determined the within- and between-subjects stability per time duration. For the within-subjects stability, we individually correlated (Pearson) the TMIF per time duration (i.e., first minute, first 3, 5, ...) with the TMIF of the entire recording per subject. This resulted in a single correlation coefficient for each participant, frequency band and time duration. Next, all correlations were plotted as a function of recording time, and we determined the knee point of the curve on the average across all frequency bands. As such, we gained insight into the amount of data required for a participant's TMIF to become stable (i.e., when there is not much change in an individual's TMIF).

\textbf{Between-subjects stability. }For the between-subjects stability, we calculated each participant's mean MI of the TMIF (integration window 0-400 ms) per time duration (1, 3, 5,... minutes) and the entire recording. This resulted in a single datapoint per participant, frequency band and time duration (i.e., mean MI for a certain duration length x mean MI entire duration). Subsequently, we calculated the correlation coefficient (Pearson's R) between the mean MI for certain time duration and frequency band with the entire recording over participants on the group level, resulting in a single correlation coefficient per time duration and frequency band. We plotted the correlations as a function of recording time and determined the knee point of the curve on the average across all frequency bands. With this analysis, we investigated the amount of data required for a participant's relative (i.e., compared to other participants) strength of tracking to become stable (i.e., from which point on a participant's relative neural tracking compared to other participants is no longer expected to change).

\subsubsection*{Split-half reliability}
Finally, we report a traditional split-half reliability metric with non-overlapping parts of the recording. We split the EEG recording into two equal parts, i.e., the first 12.5 minutes and the second 12.5 minutes, and computed the TMIFs for each half and each frequency band individually. Next, we computed the mean MI value of the TMIF (0-400 ms) for the first and the second half of the recording per participant individually. Subsequently, we calculated the correlation coefficient (Pearson's R) between the first and second half of the recording over participants on the group level (Pearson's R) for IWA and controls separately.  

\subsection*{Data availability statement}
We shared our neural tracking outcomes (i.e., the TMIFs) on the Open Science Framework: \url{https://osf.io/nkmfa/}. Note that our ethical approval does not permit public archiving of raw neuroimaging data, but raw EEG data can be made available upon request and if the GDPR-related conditions are met. 

\section{Results}
\subsection{Distinguishing individuals with aphasia from healthy controls}
\begin{figure}[H]
\centering
\includegraphics[width=1\textwidth]{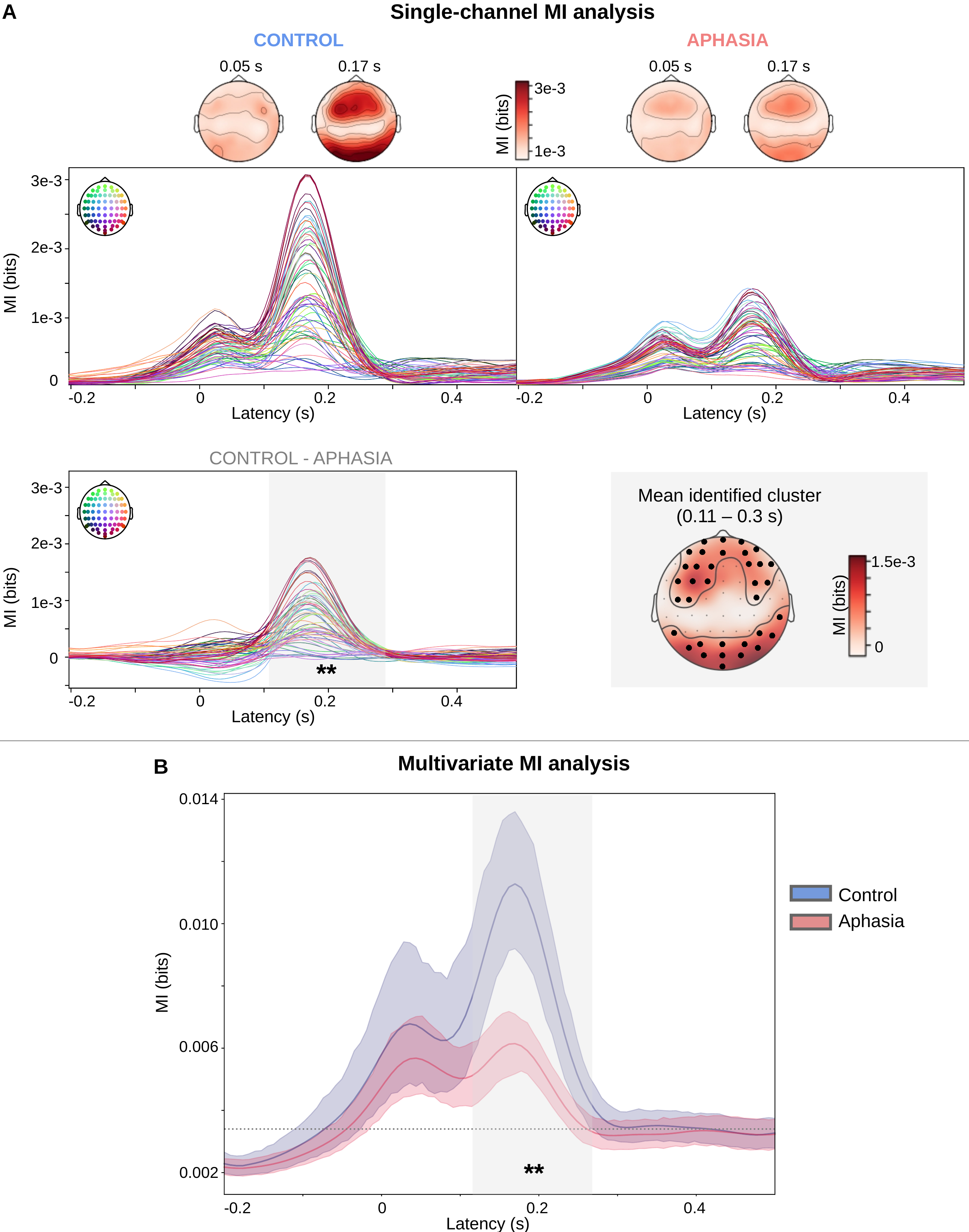}
\caption[Broadband]{\textbf{Broadband analysis.} \textbf{A. }The average single-channel TMIF for the control and the aphasia group separately, with topoplots at the first and second peak (0.05 and 0.17~s). The spatio-temporal cluster-based permutation test investigated the difference between the control and aphasia group (control - aphasia) and identified a cluster (below threshold p<0.05) with the largest group difference, centered around the second peak. Brain latencies belonging to the cluster are marked in a shaded gray area, the channels belonging to the cluster are indicated with a black dot on the topoplot. \textbf{B. } The group average TMIF, for both groups separately. The shaded, colored areas indicate the 95\% confidence interval. The shaded gray area indicates the cluster with largest group difference (threshold p<0.05), identified using a temporal cluster-based permutation test. ** = p<0.01}
\label{fig:broadband}
\end{figure}

We investigated whether neural envelope tracking is altered in IWA compared to healthy controls. First, we studied the effect in the broadband frequency range (0.5-49~Hz). For the single-channel MI analysis, providing both temporal and spatial information, we found decreased neural envelope tracking for IWA compared to healthy controls (Fig. \ref{fig:broadband}A). A spatio-temporal cluster-based permutation test identified a cluster comprising a large group of fronto-central, parietal and posterior channels (N = 43 channels) from 0.11 s to 0.3 s (p=0.004), centered around the second peak. The multivariate MI analysis, which combines information from multiple channels, confirmed these results: a temporal cluster-based permutation test identified a cluster between 0.11 s to 0.26 s in which IWA displayed a decreased response (p = 0.005) (Fig. \ref{fig:broadband}B).

\begin{figure}[H]
\centering
\includegraphics[width=1\textwidth]{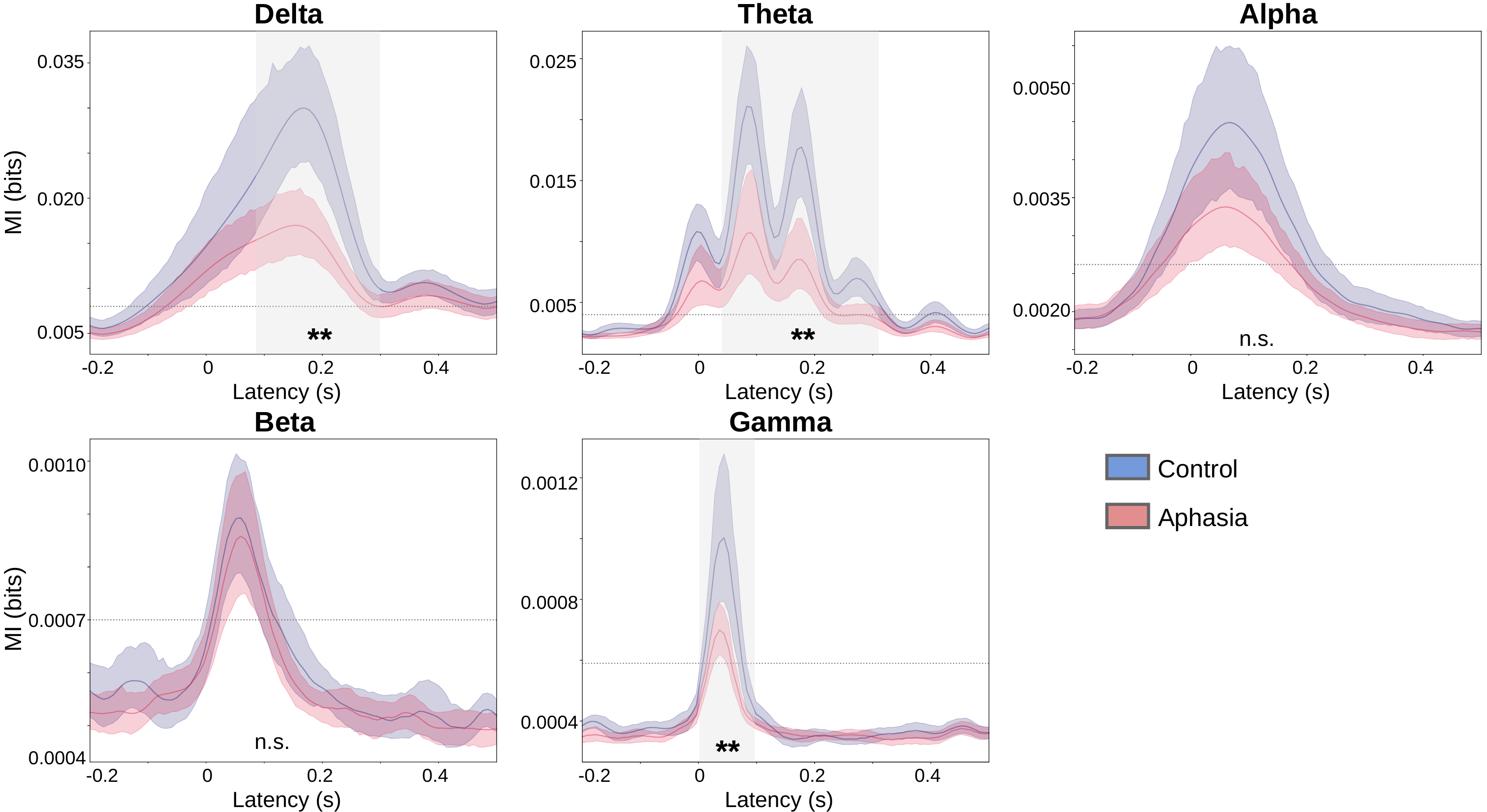}
\caption[Frequencies]{\textbf{Frequency-specific analysis.} Group average TMIF's visualized per frequency band, with colored shaded areas indicating the 95\% confidence interval. Shaded, gray areas indicate clusters with largest group difference (below threshold p<0.05) identified using temporal cluster-based permutation tests. ** = p<0.01}
\label{fig:freq}
\end{figure}

We further investigated the neural response in narrow frequency bands. We focused on the multivariate TMIF as it is a statistically more robust method compared to the single-channel TMIF, and we used those features as input to our SVM classifier in the subsequent section. The single-channel TMIFs for all frequency bands are provided in the Supplementary materials. We generally observed decreased neural envelope tracking for IWA compared to healthy controls (Fig. \ref{fig:freq}). Temporal cluster-based permutation tests identified clusters below threshold p<0.05 for delta (0.1 to 0.30s, p=0.003), theta (0.04 to 0.27s, p=0.005) and gamma (0.01 to 0.1s, p = 0.004) band. No clusters exceeding the p<0.05 threshold were detected for the alpha and beta bands. These results are confirmed in the single-channel MI analysis, where spatio-temporal cluster-based permutation tests identified clusters for delta, theta and gamma band for a large group of fronto-central, parietal and posterior channels (visualizations and statistics provided in the Supplementary materials).

\subsection{Support Vector Machine classification}
\begin{figure}[H]

\includegraphics[width=1\textwidth]{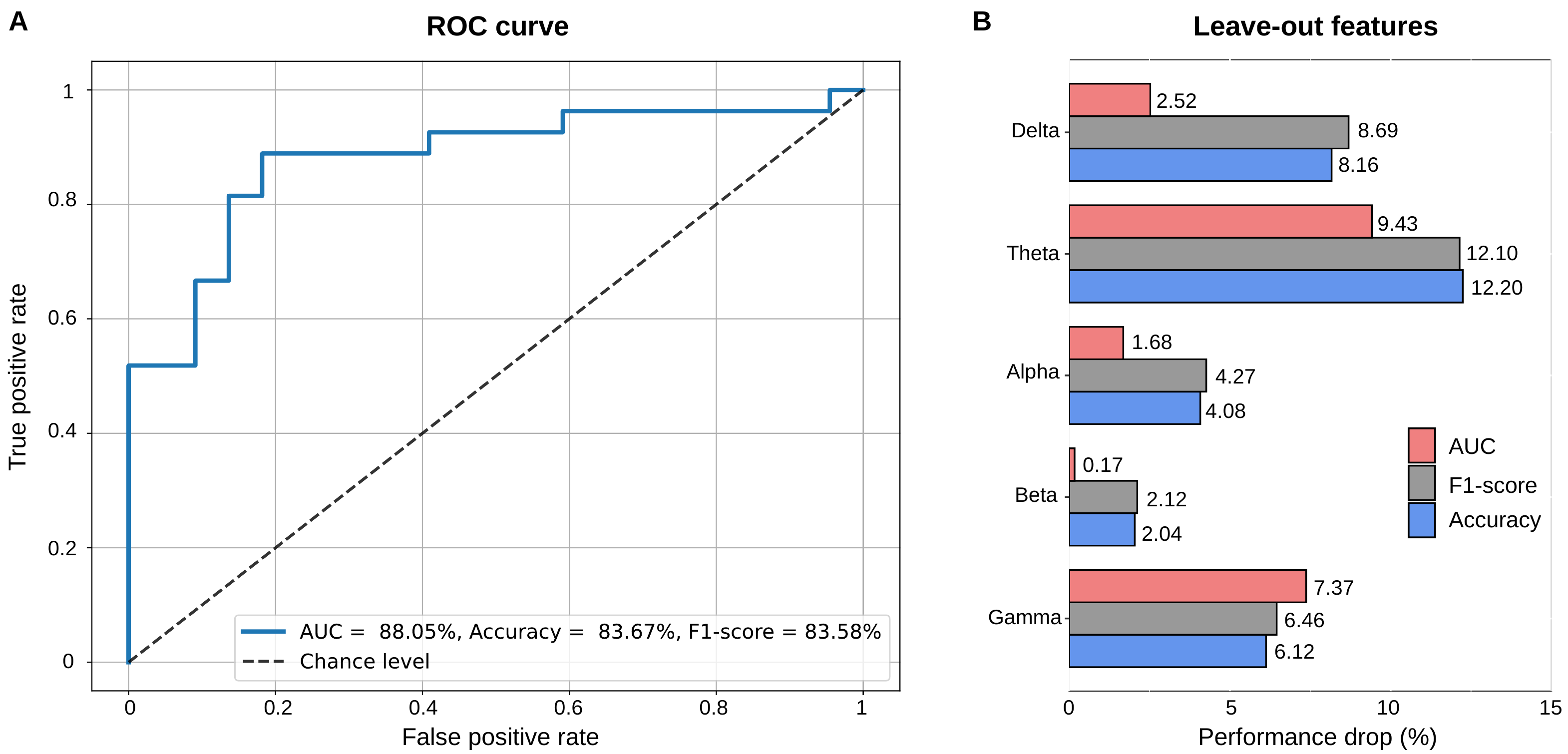}
\caption[Recording]{\textbf{Result of the SVM classifier. }\textbf{A.} Receiver operating characteristic curve (ROC). \textbf{B. }Relative feature importances. Drop in performance is visualized after leaving out the corresponding feature (i.e., frequency band). }
\label{fig:classifier}
\end{figure}

Next, we investigated whether we could detect aphasia based on neural envelope tracking measures in the individual frequency bands. To this end, we used an SVM to classify participants as belonging to the aphasia or the healthy control group via leave-one-out cross-validation. We used the TMIFs in our five frequency bands of interest and age as input features to the model. The SVM successfully classified participants belonging to either group with an accuracy of 83.67\%, an F1-score of 83.58\% and an AUC of 88.05\%. The SVM had a sensitivity of 88.89\% and a specificity of 77.27\% for aphasia. Fig. \ref{fig:classifier}A displays the ROC curve. 

To obtain a measure of relative frequency band contribution, we iteratively left out a frequency band and trained the SVM with the remaining features. For each left-out frequency band, we calculated the performance drop. As assessed with accuracy and F1-score metrics, theta, followed by delta, gamma, alpha and beta caused the largest drop in performance (see Fig. \ref{fig:classifier}B). When estimated with AUC, the accuracy was still the highest for theta, followed by gamma, delta, alpha and beta. This confirmed our group comparison analyses: delta, theta and gamma band are the most relevant, discriminating features.

\subsection{Recording length}
\begin{figure}[H]
\centering
\includegraphics[width=1\textwidth]{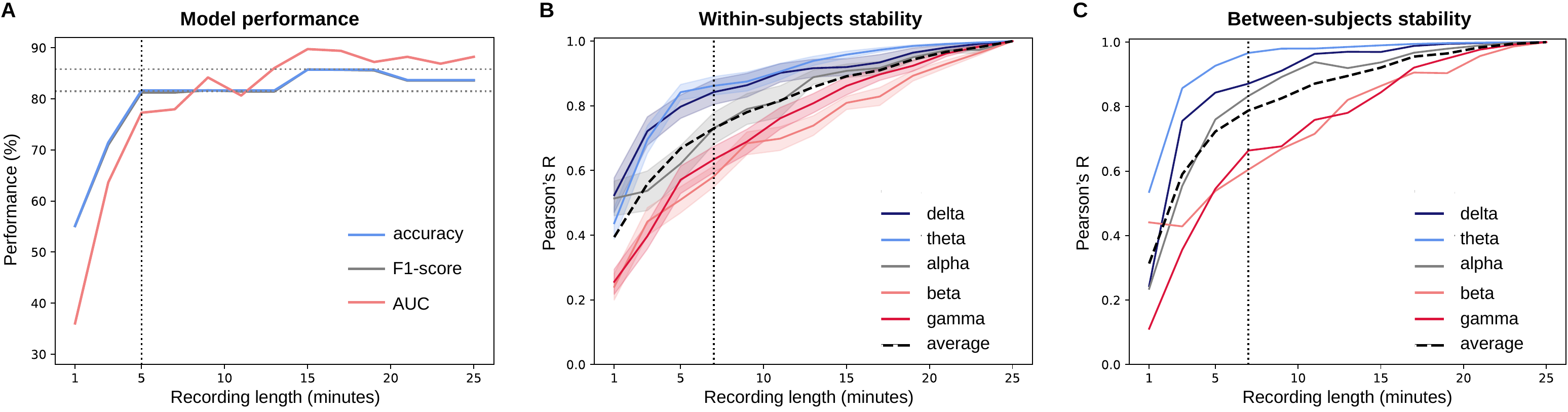}
\caption[Recording]{\textbf{Recording Length. } \textbf{A.} Performance (accuracy, F1-score, AUC) of the SVM classifier plotted as a function of time. \textbf{B. }Within-subjects stability for all five frequency bands and the average across frequency bands (black line). Shaded areas indicate the standard error. \textbf{C. }Between-subjects stability for all five frequency bands and the average across frequency bands (black line). The knee point of all panels is indicated with a vertical dotted line (based on the average for panels \textbf{B} and \textbf{C}).}
\label{fig:recording}
\end{figure}

We further investigated how much data the neural envelope tracking technique requires for robust and stable results (Fig. \ref{fig:recording}). With only one minute of recording time, the SVM obtained classification accuracy close to chance-level (55\%). Yet, from 5 minutes on, the SVM reached an accuracy of 81.63\%, and performance fluctuated between 81.63\% and 85.71\% for the remaining part of the recording. In practice, this corresponds to one less or one additional correctly classified participant with respect to the full recording (Fig. \ref{fig:classifier}A). The knee point of the curve was identified at 5 minutes of recording length. From 9 minutes on, the SVM converged to an AUC of 80\%. However, compared to the entire recording length (AUC=88.05\%), its full potential is reached from 13 minutes on (AUC robustly crossed 85\%, with a maximum of 89.73\% at 15 minutes). The F1-score mostly overlapped with the accuracy and never differed more than 0.44\% . The SVM performance is plotted as a function of recording length, displayed in Fig. \ref{fig:recording}A. 

The within-and between-subjects stability is plotted as a function of recording time in Fig. \ref{fig:recording}B and \ref{fig:recording}C. Highest within- and between-subjects correlations were observed for the low-frequency bands, namely delta and theta. Taking the average of all frequency bands, we identified the curve's knee point at 7 minutes of recording length (see black dotted lines). The within-subjects stability (Fig. 1 \ref{fig:recording}B) had an average correlation of R=0.73, and the between-subjects stability (Fig. \ref{fig:recording}C) had an average correlation of R=0.79 at the knee point of the curve.

\subsection{Reliability of neural envelope tracking}
Finally, we calculated the split-half reliability of neural envelope tracking. Table \ref{table:splithalf} provides the correlations and statistics. We generally found higher correlations in the lower frequency bands (delta and theta). Correlations were comparable between IWA and the control group; the 95\% confidence intervals overlapped for each frequency band, and post-hoc Fisher z-tests, performed using the 'cocor' package in Rstudio \citep{Diedenhofen2015}, revealed no difference in correlation strength between both groups.
\begin{table}[h]
\caption{Split-half reliability.}
\centering
\resizebox{\textwidth}{!}
{\begin{tabular}{ccccccccccc}
\toprule
\multicolumn{1}{c}{} & \multicolumn{2}{c}{\textbf{delta}} & \multicolumn{2}{c}{\textbf{theta}} & \multicolumn{2}{c}{\textbf{alpha}} & \multicolumn{2}{c}{\textbf{beta}} & \multicolumn{2}{c}{\textbf{gamma}}\\
\cmidrule(rl){2-3} \cmidrule(rl){4-5} \cmidrule(rl){6-7} \cmidrule(rl){8-9} \cmidrule(rl){10-11}
\textbf & {IWA} & {C} & {IWA} & {C} & {IWA} & {C} & {IWA} & {C}  & {IWA} & {C}\\
\midrule
\\ Pearson's R & 0.85 & 0.91 & 0.96 & 0.92 & 0.75 & 0.88 & 0.61 & 0.62 & 0.62 & 0.73\\
\\
CI & [0.68 ; 0.93] & [0.78 ; 0.96] & [0.91 ; 0.98] & [0.82 ; 0.96] & [0.52 ; 0.88] & [0.72 ; 0.95] & [0.31 ; 0.81] & [0.27 ; 0.83] & [0.31 ; 0.81] & [0.43 ; 0.88]\\
\\
p-value & <0.001 & <0.001 & <0.001 & <0.001 & <0.001 & <0.001 & 0.006 & 0.02 & 0.007 & 0.002\\
\\
Fisher's z (p-value) & \multicolumn{2}{c}{z=0.85 (1)} & \multicolumn{2}{c}{z=-0.94 (1)}& \multicolumn{2}{c}{z=1.24 (1)}& \multicolumn{2}{c}{z=0.04 (1)}& \multicolumn{2}{c}{z=0.66 (1)}\\
\\
\bottomrule
\end{tabular}}
\\
\begin{flushleft}
\footnotesize{CI=95\% confidence interval; C=Control group. Fisher z-test comparing controls - IWA. All p-values are corrected for multiple comparisons.}
\end{flushleft}
\label{table:splithalf}
\end{table}

\newpage

\section{Discussion}
We conducted an in-depth study on neural envelope tracking of natural speech in post-stroke aphasia. First, we found that IWA display decreased neural envelope tracking compared to heatlhy controls for a broadband frequency range. Second, frequency-specific analyses indicated that group differences are most prominent in the delta, theta and gamma frequency ranges. Third, the suitability of neural envelope tracking measures as a biomarker for post-stroke aphasia was demonstrated using an SVM classifier which yielded high accuracy (84\%, AUC 88\%). Finally, we showed that an assessment based on neural envelope tracking could be obtained in a time-efficient (5 minutes of EEG recording) and highly reliable manner.

\subsection{Individuals with aphasia display decreased neural envelope tracking}
\subsubsection*{Broadband frequency analysis}
Neural envelope tracking at broadband is decreased in IWA compared to healthy controls. The single-channel TMIF analysis revealed a cluster at neural response latencies centered around the second peak in the TMIF comprising a large group of fronto-central, temporal and parieto-occipital channels (see Fig. \ref{fig:broadband}A). The multivariate TMIF confirmed this result: a temporal cluster comprising brain latencies surrounding the second peak in the TMIF (Fig. \ref{fig:broadband}B) was identified. A recent neural tracking study showed that the second peak emerges when speech is comprehensible and diminishes when it is not understood. By contrast, the first peak displayed a prominent response when speech was incomprehensible \citep{Verschueren2022}. Thus, the second peak we observed here is most likely related to speech understanding, while the first peak is likely more implicated in acoustically processing the signal. Therefore, it is not surprising that in IWA, where language understanding is impaired, the second peak in the TMIF is decreased compared to healthy controls.

\subsubsection*{Frequency-specific analysis}
We further investigated neural envelope tracking in narrow frequency bands. Our findings revealed a decrease in tracking for IWA compared to healthy controls in the low-frequency bands (delta and theta, see Fig. \ref{fig:freq}A and \ref{fig:freq}B), which are crucial for speech understanding \citep{Vanthornhout2018}. The delta band encodes sentences, phrases and words \citep{Kaufeld2020, Keitel2018}, while theta band tracks the syllable rate of the stimulus \citep{Etard2019, Lizarazu2019}. Neural tracking in the low-frequency bands drops when these linguistic units become incomprehensible \citep{Kaufeld2020, Keitel2018, Xu2023}. Atypical neural tracking of the low-frequency temporal envelope has been reported in several clinical populations, including individuals with primary progressive aphasia \citep{Dial2021} and dyslexia \citep{DiLiberto2017, Lizarazu2021, Mandke2022}. In the case of dyslexia, which is characterized by phonological processing difficulties, alterations in low-frequency envelope tracking are believed to reflect an atypical sampling mechanism that affects faster modulations at the phoneme and grapheme level \citep{Mandke2022}. These findings in healthy and clinical populations demonstrate the potential of neural tracking measures of the low-frequency envelope as a biomarker for language impairments.

Fewer studies investigated the role of high-frequency neural envelope tracking. Some studies suggest a role for alpha and beta in attention and auditory-motor coupling \citep{Wostmann2017, Fujioka2015}, and for the gamma band in encoding phonetic features \citep{Hyafil2015, Giraud2012, Gross2013}. Our study found no group differences in the alpha and beta bands. However, individuals with aphasia displayed decreased neural envelope tracking in the gamma band. The neural response in the gamma band was characterized by an early response peak (Fig. \ref{fig:freq}) and a group difference present in the right hemisphere (see Supplementary Fig. \ref{fig:gamma_sc}). This early response latency in the gamma band aligns with the idea of a linear phase property, where the neural response delay in higher frequency bands is shorter \citep{Zou2021}. A somewhat similar neural response pattern characterized by an early response latency and a right hemisphere bias in the high gamma band (>70~Hz) was also reported by \citet{Kulasingham2020}. The gamma band has been of particular interest in dyslexia research, with several studies reporting alterations in gamma band activity. However, most studies have focused on phase-locking and phase coherence in response to amplitude-modulated noise (for a review, see \citep{Lizarazu2021}), while few have investigated gamma band neural envelope tracking during natural speech processing \citep{Mandke2022}. To gain a better understanding of (low-)gamma band neural envelope tracking of natural speech, which we have shown to demonstrate robust group differences and contribute significantly to detecting aphasia (as depicted in Fig. \ref{fig:classifier}B), future research should aim to further investigate its implications for speech understanding and language impairments. 

In line with the idea that individual frequency bands are involved in different speech processes, exploratory analysis revealed low to moderate positive and negative correlations between neural tracking in individual frequency bands (see Supplementary Table \ref{table:corr}). This suggests that a participant with high neural tracking in one frequency band may not necessarily display high neural tracking in other frequency bands. In contrast, a broadband frequency analysis shows high redundancy compared to the delta band (R=0.79 for IWA, R=0.92 for controls), which can be attributed to the fact that most power in the EEG and the envelope is concentrated in the lowest frequencies. This highlights the relevance of conducting frequency-specific analyses. We believe that the use of frequency-specific features to train the SVM favored good classification results, as discussed in the next section. Future research should investigate whether the neural response to these frequency bands may capture specific language deficits.

\subsection{High accuracy detection of post-stroke aphasia}
We assessed the suitability of the neural tracking technique to detect post-stroke aphasia using an SVM classifier with the TMIFs computed in the individual frequency bands as input to the model. The SVM robustly detected aphasia, with an accuracy of 83.67\%, an F1-score of 83.58\% and an AUC of 88.05\%. The ROC curve, plotting the true positive as a function of false positive aphasia classification, is depicted in Fig. \ref{fig:classifier}A. The relative contribution of individual frequency bands for detecting aphasia at the individual level confirmed our group comparison results: delta, theta and gamma band neural tracking were most predictive for capturing aphasia (see Fig. \ref{fig:classifier}B).

These performance outcomes of the SVM can be interpreted against behavioral assessment. As described in the Methods section, 7 out of 27 IWA (i.e., 26\%) did not score below the cut-off threshold on either of the two diagnostic language tests for aphasia administered during the study \cite{ElHachioui2017,nbt}. Nevertheless, these subjects had extended language deficit documentation and followed speech-language therapy at the time of participation. Although a more extensive screening for aphasia could have identified a language deficit, this finding highlights the challenge of detecting aphasia in the chronic phase following a stroke. While further investigation is required, the higher detection accuracy of the EEG-based neural tracking classification suggests that it may be more sensitive than behavioral screening tests for capturing subtle language problems in individuals with aphasia.

Nonetheless, comparing the SVM classification accuracy to behavioral assessment is rather difficult, as the underlying tested language skills are different. Standardized aphasia tests use isolated sounds, phonemes, words or short sentences, questioning the ecological validity of such tasks \citep{Hamilton2018}. Consequently, research reports a discrepancy between common test outcomes and everyday life speech assessments \citep{Lesser1995,Kim2022, Stark2021, Wallace2013}, and cognitive problems can bias the test result \citep{Fonseca2019, Rohde2018}. While there exist behavioral natural speech assessments \citep{Armstrong2000}, they are only limitedly applied in practice due to the high workload (time-intensive) and a lack of knowledge of natural speech analyses \citep{Bryant2019, Stark2021}. Novel automatic speech recognition and natural language processing techniques \citep{Dalton2022, Jamal2017, Le2018} may provide a solution in the future. The neural tracking technique directly addresses the limitation of low ecological validity from which behavioral tests suffer. 

\subsection{Assessing time-efficiency, stability and reliability}
We investigated how much data neural envelope tracking requires to detect aphasia accurately and yield reliable results. We assessed SVM classification performance as a function of recording time, as shown in Fig. \ref{fig:recording}A. Our findings indicate that high-accuracy detection can be achieved with just 5 minutes of recording time (accuracy of 81.63\%). However, extending the recording duration to 13 minutes can provide additional benefits in terms of the AUC, with a robust increase above 85\%, and a maximum AUC of 89.73\% achieved at 15 minutes. In summary, our results demonstrate that neural envelope tracking can effectively detect aphasia in a time-efficient manner, consistent with prior recommendations that language assessments in aphasia should not exceed 15 minutes to avoid fatigue and cognitive/attentional challenges \citep{ElHachioui2014}. Our findings have important implications for potential clinical applications of neural envelope tracking in aphasia.

We further investigated the amount of data necessary for our frequency band features to achieve stability. Our within- and between-subjects stability analysis revealed that 7 minutes recording length is sufficient for the TMIF of individual subjects and at the group-level to resemble the TMIF from the entire recording  (see Fig. \ref{fig:recording}B and \ref{fig:recording}C). These findings were consistent within both groups, with 7 minutes being the minimum recording length required (as illustrated in Supplementary Fig. \ref{fig:stab_groups}). Notably, we observed that lower frequency bands (delta and theta) converge more rapidly compared to higher frequency bands. Within 3-5 minutes, stability correlations for these bands were relatively high, robustly crossing R=0.80. In contrast, higher frequency bands require a longer time to converge and exhibit a more linear slope compared to the delta and theta bands. These less stable results and longer minimal recording length for the higher frequency bands can be attributed to their lower signal-to-noise ratio.

To summarize, our results indicate that 5-7 minutes of recording time are sufficient for assessing neural envelope tracking at low-frequency ranges, which reflect higher-level linguistic processes and speech understanding. However, a more comprehensive evaluation that includes higher frequency bands, which can provide minor additional benefits, requires a slightly longer recording duration (>13 minutes). These findings are consistent with previous research in healthy participants, which suggests that low-frequency neural tracking requires approximately 3-10 minutes of recording time for robust outcomes \citep{Desai2023, DiLiberto2017, Mesik2022}. Our study contributes an innovative approach by defining the minimal recording length required to detect language impairments at the individual level and suggests that the recording duration for future studies in individuals with language impairments may depend on the specific research question being addressed.

Previous studies on aphasia using ERPs have suggested the potential of this approach for clinical diagnosis, but without reporting on its reliability \citep{Cocquyt2020}. However, evaluating the reliability of test results is crucial to determine the usefulness of capturing individual language impairments. In this study, we assessed the reliability of neural envelope tracking using split-half reliability metrics. The results demonstrate strong correlations between both halves, particularly in the delta and theta bands (Table \ref{table:splithalf}). Our findings are consistent with previous research reporting a correlation of R=0.89 for delta and R=0.82 for theta across two stories in a cohort with language impairments caused by a neurodegenerative disorder \citep{Dial2021}. Yet, reliability measures in our study were generally lower for higher frequency bands. As mentioned earlier, these bands require more data to converge and have a lower signal-to-noise ratio. It is worth noting that our reliability measure may be affected by fatigue. Thus, future studies should examine the generalizability of the results across stories and speakers at different sessions (i.e., test-retest) to further investigate the reliability of neural tracking for applications in aphasia.

\subsection{Limitations and future directions}
Our study demonstrates that neural envelope tracking is a reliable and accurate method for detecting language impairments in aphasia. However, our current approach does not provide information on the specific language profile of the patient (i.e., which underlying language component, e.g., auditory, phonetic, semantic,... is affected). Investigating these deficits would require a larger sample size with a more uniform spread of aphasia severity levels. In future research, we suggest exploring whether neural tracking in specific frequency bands can cluster different language profiles in aphasia. In addition, recent studies investigated the neural response to speech representations beyond the temporal envelope. For example, it has been shown that linguistic speech representations at phoneme and word level can improve the model's fit to the EEG \citep{DiLiberto2015, Gillis2021} and can provide complementary information on speech processing \citep{Verschueren2022, GillisKries}. Future research should (1) examine whether incorporating these linguistic speech representations can enhance aphasia detection and inform on specific language deficits and (2) assess the reliability and robustness of these features, which is currently lacking in the literature.

Several other open questions must be addressed before neural tracking can be applied in clinical settings. Firstly, neural tracking must be applied to IWA in the acute stage after stroke. This work considered the chronic stage only since it is characterized by a more stable language profile \citep{Johnson2019}. Secondly, the present study distinguished IWA and healthy controls only. If neural tracking is to be used for screening aphasia in the acute stage post-stroke, a clear dissociation between stroke patients with and without aphasia is crucial. However, such dissociation is generally not considered in behavioral screening tests despite being used in the clinic on a daily basis \citep{Rohde2018}. Lastly, this study used language stimuli in the receptive domain only. Recent studies have suggested that the same analysis can also be applied to the expressive domain, i.e., speech production \citep{Perez2022}, which could open new perspectives to studying expressive language problems in IWA. 

\subsection*{Conclusion}
This study investigated neural envelope tracking of natural speech in patients with chronic post-stroke aphasia. The findings showed that individuals with aphasia exhibited reduced brain responses in the delta, theta, and gamma bands, likely reflecting decreased processing of higher-level auditory and linguistic units. The study also demonstrated the efficacy of neural tracking in capturing language impairments at the individual level in a highly reliable and time-efficient manner, which suggests its promising clinical potential as an assessment tool. Despite these positive results, several open questions remain that need to be addressed before neural tracking can be used in clinical settings. For instance, it remains unclear whether neural tracking can accurately capture specific language problems, and its effectiveness in assessing patients in the acute stage post-stroke requires future investigation. Nevertheless, our work represents a significant step towards more automatic and ecologically valid assessments of language problems in aphasia.

\section*{Acknowledgements}
The authors would like to express their heartfelt gratitude to all the participants, particularly those with aphasia and their families that supported them. We would also like to extend our thanks to Dr. Klara Schevenels for her assistance in the recruitment process, as well as the individuals that helped with the data collection: Janne Segers, Rosanne Partoens, Charlotte Rommel, Ines Robberechts, Laura Van Den Bergh, Anke Heremans, Frauke De Vis, Mouna Vanlommel, Naomi Pollet, Kaat Schroeven, Pia Reynaert and Merel Dillen. 

\section*{Funding}
Research of Pieter De Clercq was supported by the Research Foundation Flanders (FWO; PhD grant 1S40122N). Jill kries was financially supported by the Luxembourg National Research Fund (FNR; AFR-PhD project reference 13513810). Research of Jonas Vanthornhout was supported by FWO (postdoctoral grant: 1290821N). The presented study further received funding from the European Research Council (ERC) under the European Union’s Horizon 2020 research and innovation programme (Tom Francart; grant agreement No. 637424), and by the FWO grant No. G0D8520N.

\section*{Competing interests}
The authors declare no conflicts of interest, financial or otherwise.

\setcounter{table}{0}
\setcounter{figure}{0}
\newpage
\section*{Supplementary material}
\subsection*{Channel Selection}

\begin{figure}[H]
\renewcommand{\figurename}{Supplementary Fig.}
\centering
\includegraphics[width=0.4\textwidth]{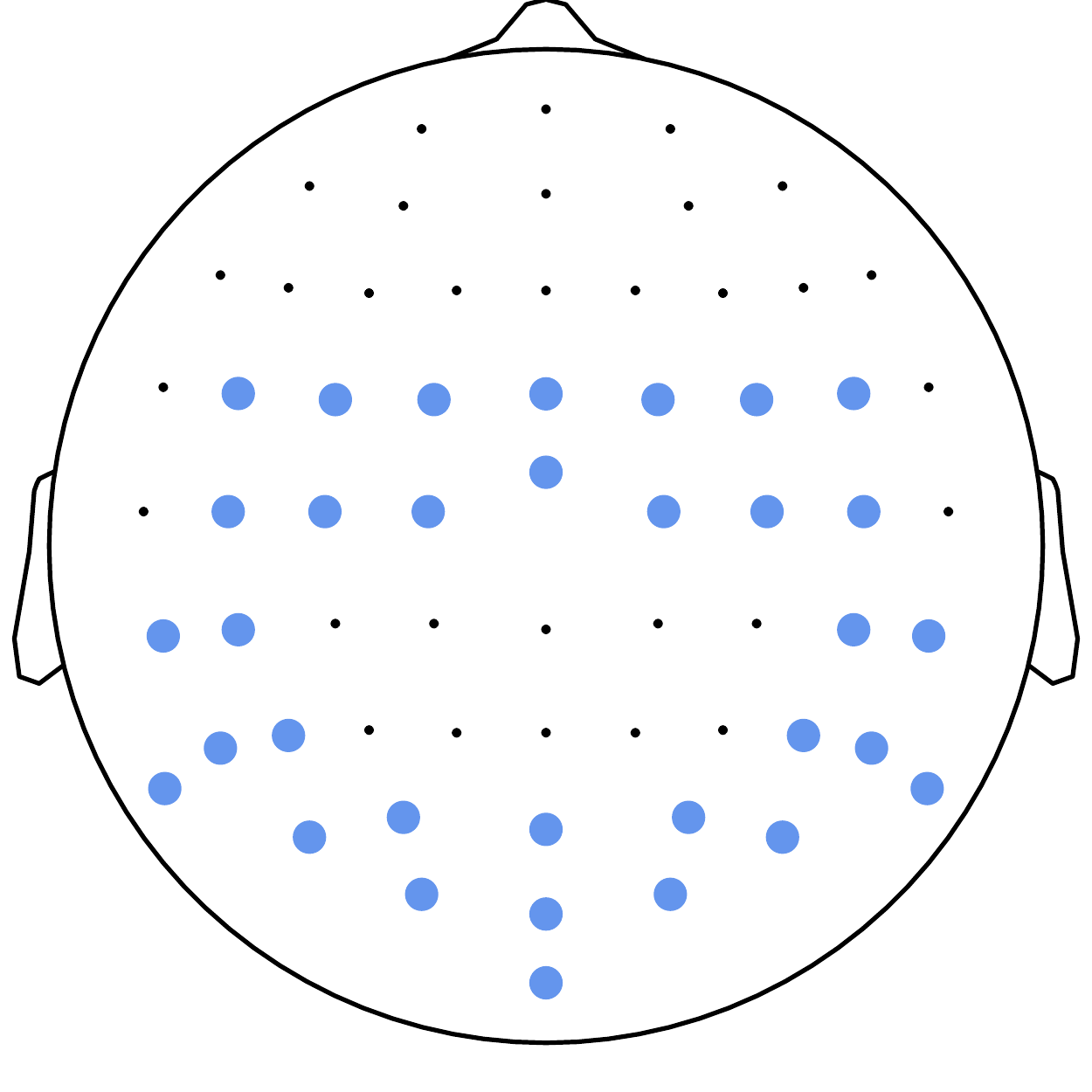}
\caption[cs]{\textbf{Channel selection. }} 
\label{fig:ch_s}
\end{figure}
\newpage
\subsection*{Significance level of neural envelope tracking}
\begin{figure}[H]
\renewcommand{\figurename}{Supplementary Fig.}
\centering
\includegraphics[width=1\textwidth]{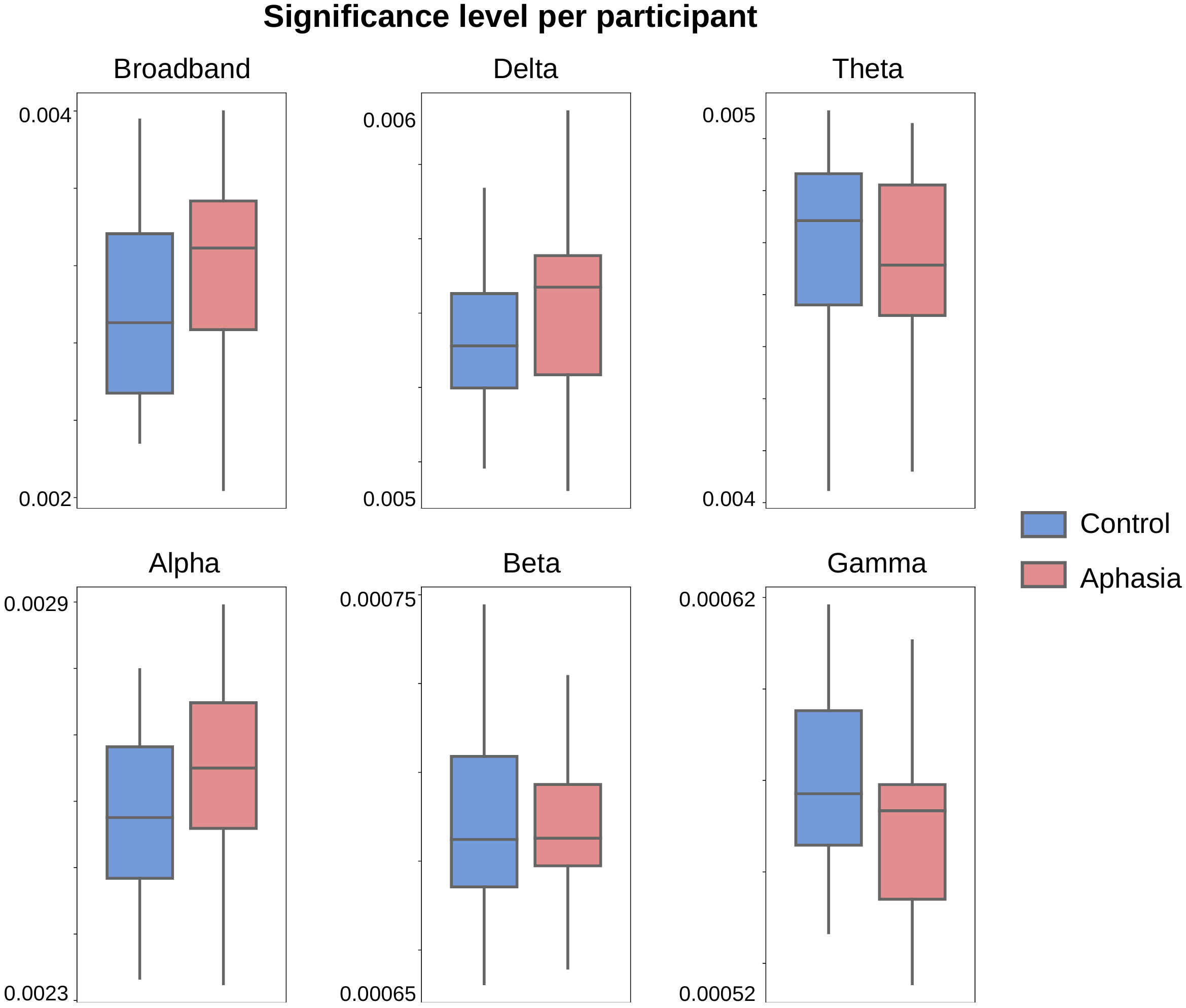}
\caption[Sigs]{\textbf{Significance level of neural tracking. }Boxes represent the 95th percentile of permutations per subject and per frequency band. There was no significant difference between groups for any frequency band. } 
\label{fig:sigs}
\end{figure}
\newpage
\subsection*{Single-channel TMIF analysis}
\subsubsection*{Delta band}
\begin{figure}[H]
\renewcommand{\figurename}{Supplementary Fig.}
\centering
\includegraphics[width=1\textwidth]{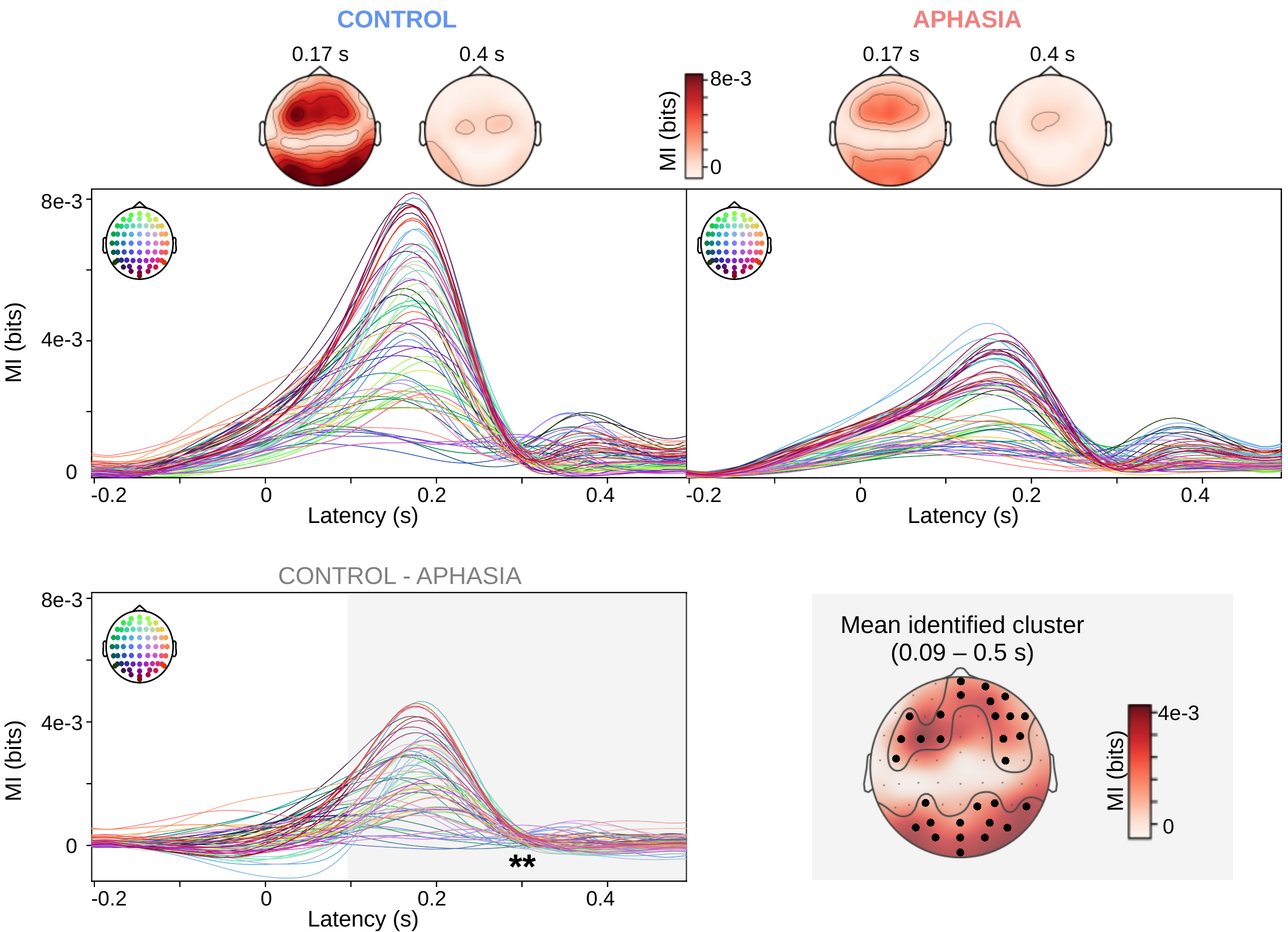}
\caption[Delta]{\textbf{Delta band analysis.} The average single-channel TMIF in delta band for the control and the aphasia group separately, with topoplots at indicated brain latencies. The spatio-temporal cluster-based permutation test investigated the difference between the control and aphasia group (control
- aphasia) and identified a cluster (below threshold p<0.05) with the largest group difference. Brain latencies belonging to the cluster are marked in a shaded gray area, the
channels belonging to the cluster are indicated with a black dot on the topoplot.** = p<0.01}
\label{fig:delta_sc}
\end{figure}

The single-channel TMIF analysis revealed decreased delta band envelope tracking for IWA compared to healthy controls. A spatio-temporal cluster-based permutation test identified a cluster (p=0.005) comprising a large group of bilateral fronto-central, parietal and posterior channels (N = 46 channels) and brain latencies from 0.09 s to 0.5 s. The results are depicted in Fig. \ref{fig:delta_sc}.

\newpage
\subsubsection*{Theta band}
\begin{figure}[H]
\renewcommand{\figurename}{Supplementary Fig.}
\centering
\includegraphics[width=1\textwidth]{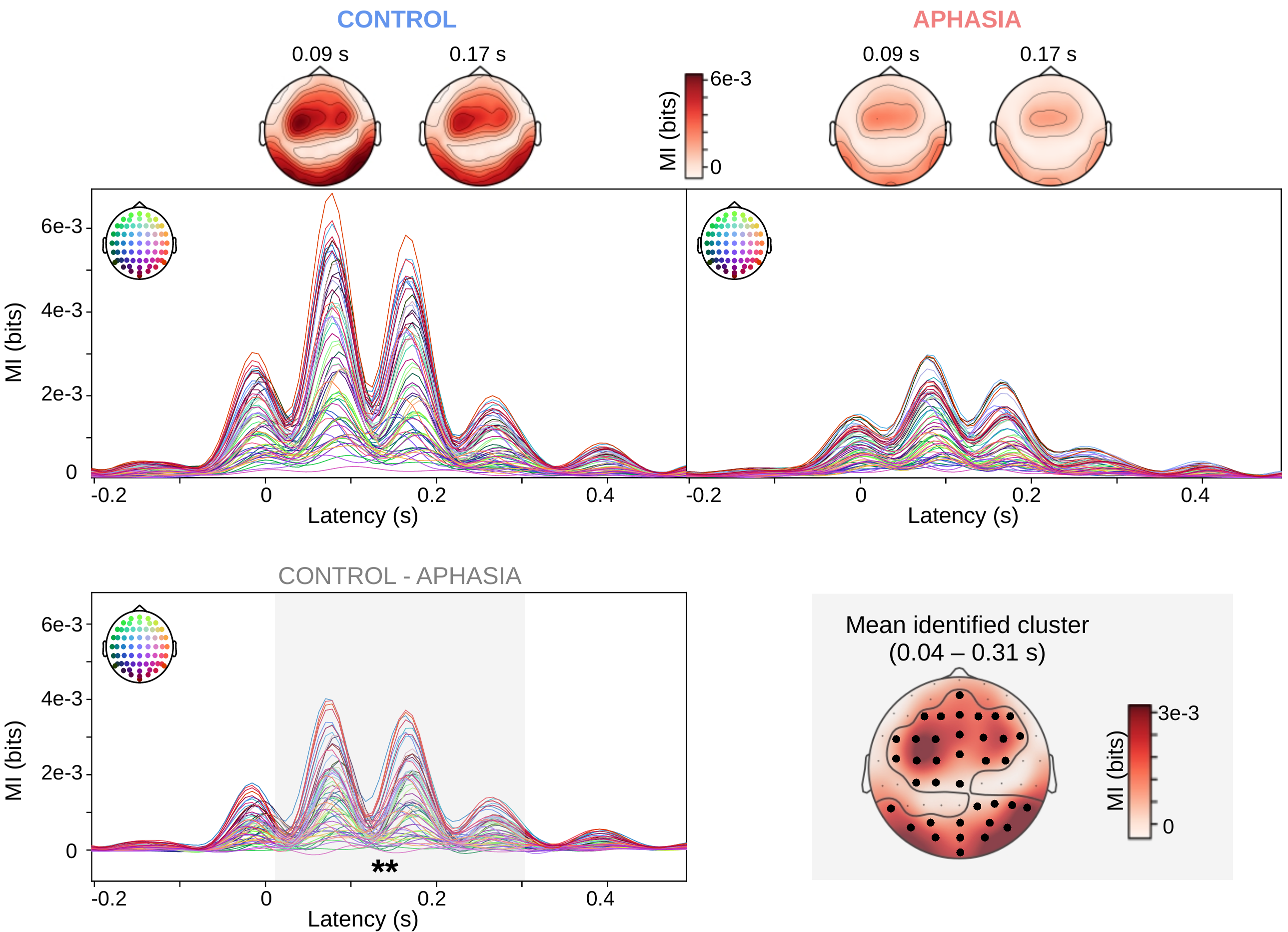}
\caption[Theta]{\textbf{Theta band analysis.} The average single-channel TMIF in theta band for the control and the aphasia group separately, with topoplots at indicated brain latencies. The spatio-temporal cluster-based permutation test investigated the difference between the control and aphasia group (control - aphasia) and identified a cluster (below threshold p<0.05) with the largest group difference. Brain latencies belonging to the cluster are marked in a shaded gray area, the
channels belonging to the cluster are indicated with a black dot on the topoplot.** = p<0.01}
\label{fig:theta_sc}
\end{figure}

For the theta band, the single-channel TMIF analysis revealed decreased envelope tracking for IWA compared to healthy controls. A spatio-temporal cluster-based permutation test identified a cluster (p=0.005) comprising a large group of bilateral fronto-central, parietal and posterior channels (N = 40 channels) and brain latencies from 0.09 s to 0.31 s. Fig. \ref{fig:theta_sc} visualizes the result.
\newpage
\subsubsection*{Alpha band}
\begin{figure}[H]
\renewcommand{\figurename}{Supplementary Fig.}
\centering
\includegraphics[width=1\textwidth]{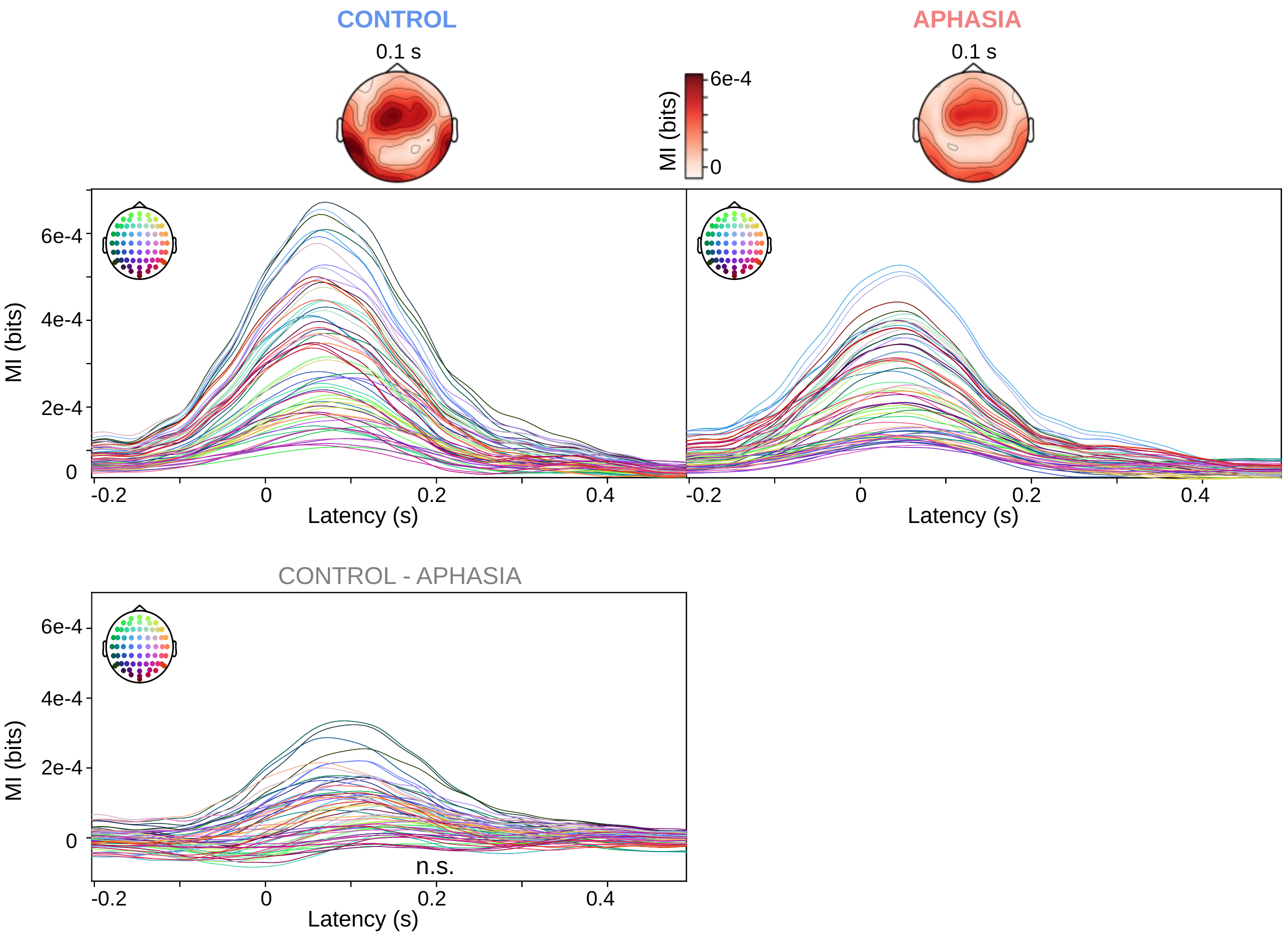}
\caption[Alpha]{\textbf{Alpha band analysis.} The average single-channel TMIF in alpha band for the control and the aphasia group separately, with topoplots at indicated brain latencies. The spatio-temporal cluster-based permutation test investigated the difference between the control and aphasia group (control - aphasia), but did not find a group difference with p-value below threshold level 0.05.}
\label{fig:alpha_sc}
\end{figure}

In the alpha band, a spatio-temporal cluster-based permutation test found no clusters exceeding p<0.05 threshold level. The group results are displayed in Fig. \ref{fig:alpha_sc}.
\newpage
\subsubsection*{Beta band}
\begin{figure}[H]
\renewcommand{\figurename}{Supplementary Fig.}
\centering
\includegraphics[width=1\textwidth]{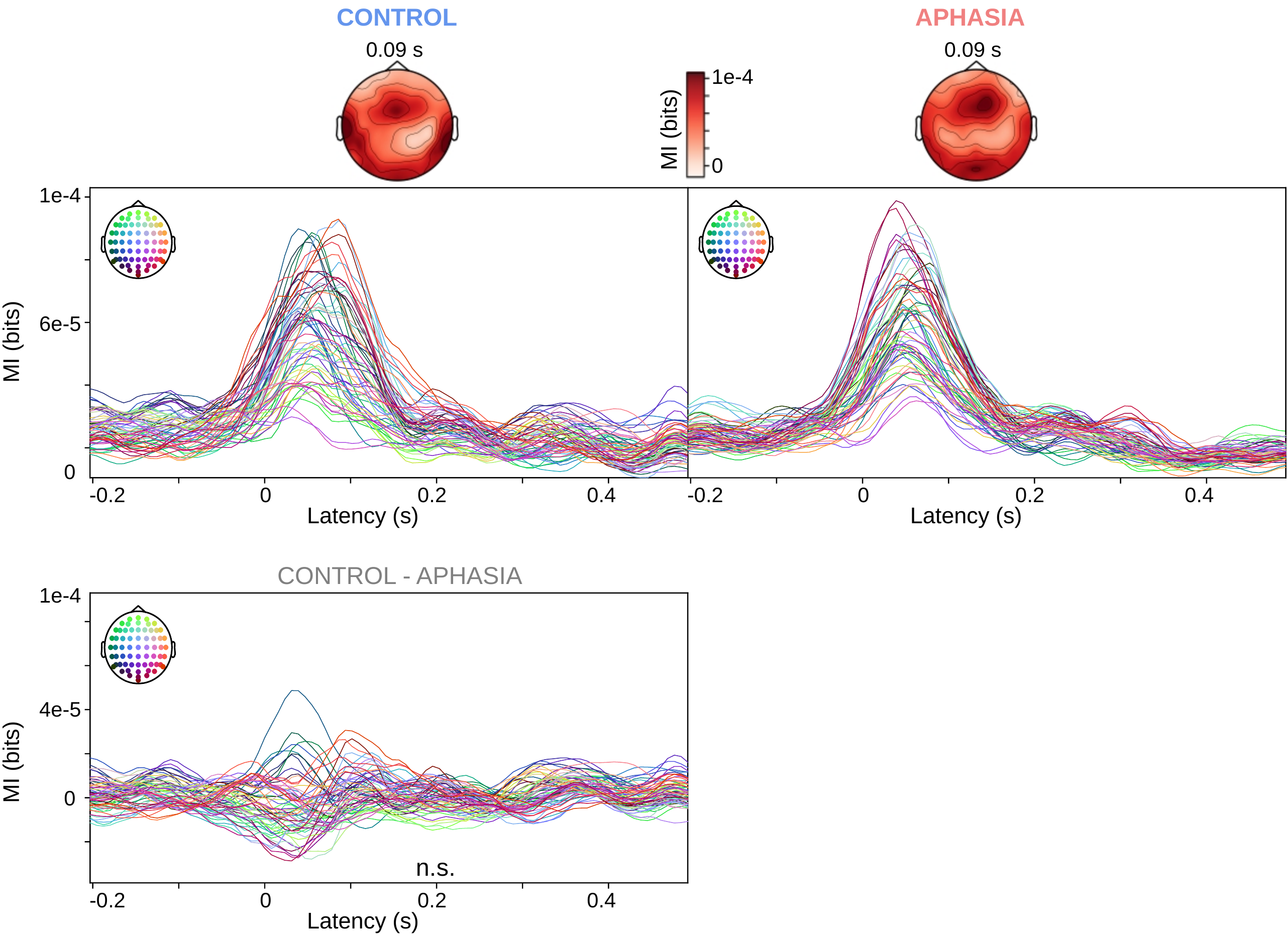}
\caption[Beta]{\textbf{Beta band analysis.} The average single-channel TMIF in beta band for the control and the aphasia group separately, with topoplots at indicated brain latencies. The spatio-temporal cluster-based permutation test investigated the difference between the control and aphasia group (control - aphasia), but did not find a group difference with p-value below threshold level 0.05.}
\label{fig:beta_sc}
\end{figure}

In the beta band, a spatio-temporal cluster-based permutation test found no clusters exceeding p<0.05 threshold level. The group results are displayed in Fig. \ref{fig:beta_sc}.
\newpage
\subsubsection*{Gamma band}
\begin{figure}[H]
\renewcommand{\figurename}{Supplementary Fig.}
\centering
\includegraphics[width=1\textwidth]{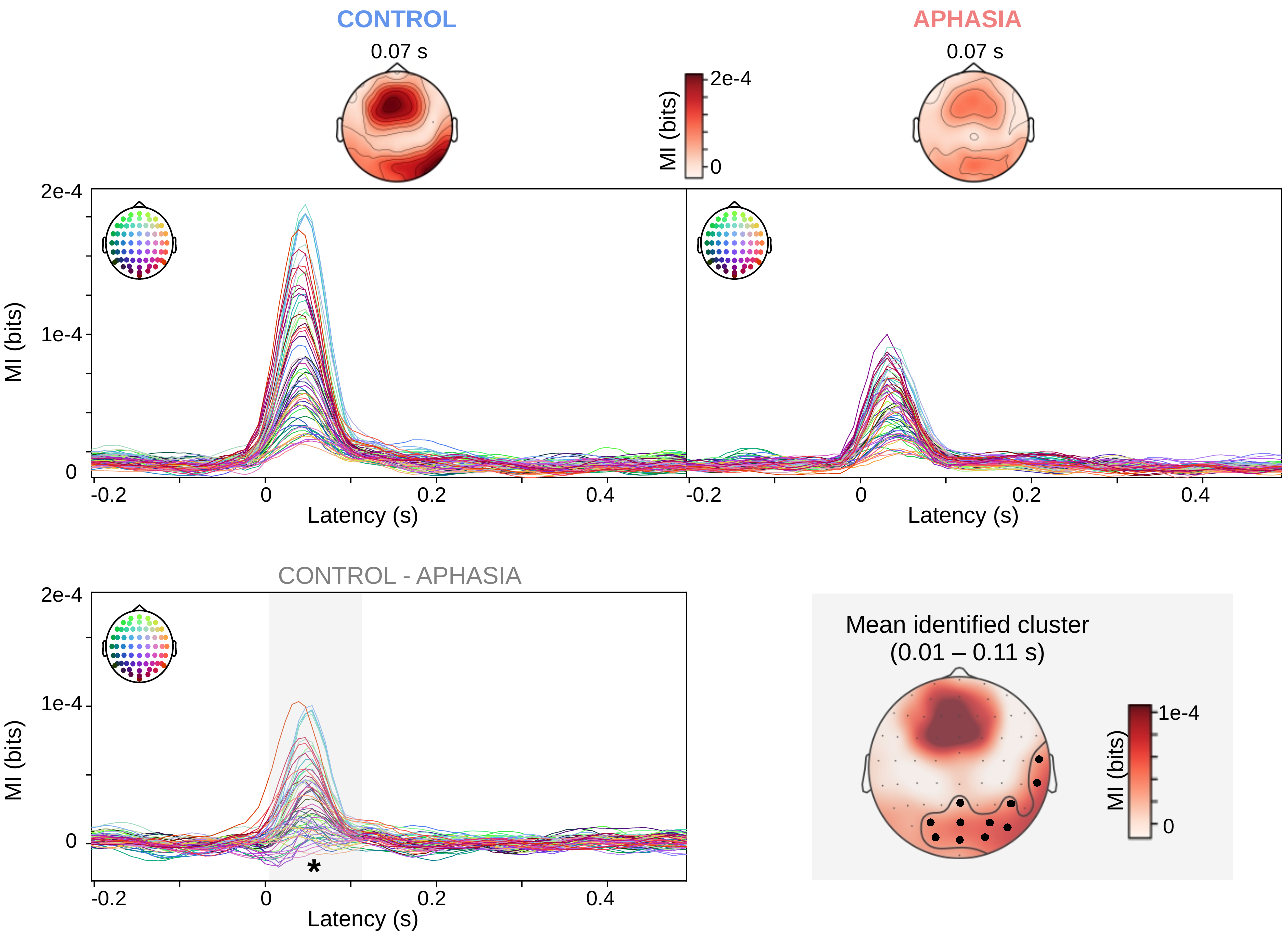}
\caption[Gamma]{\textbf{Gamma band analysis.} The average single-channel TMIF in gamma band for the control and the aphasia group separately, with topoplots at indicated brain latencies. The spatio-temporal cluster-based permutation test investigated the difference between the control and aphasia group (control - aphasia) and identified a cluster (below threshold p<0.05) with the largest group difference. Brain latencies belonging to the cluster are marked in a shaded gray area, the
channels belonging to the cluster are indicated with a black dot on the topoplot.* = p<0.05}
\label{fig:gamma_sc}
\end{figure}

Finally, IWA displayed decreased neural envelope tracking in the gamma band.  A spatio-temporal cluster-based permutation test identified a cluster (p=0.03) comprising parietal and posterior channels (N = 12 channels), primarily in the right hemisphere, and brain latencies from 0.01 s to 0.11 s. Fig. \ref{fig:gamma_sc} visualizes the result.
\newpage

\subsection*{Group-specific stability analysis}
\begin{figure}[H]
\renewcommand{\figurename}{Supplementary Fig.}
\centering
\includegraphics[width=1\textwidth]{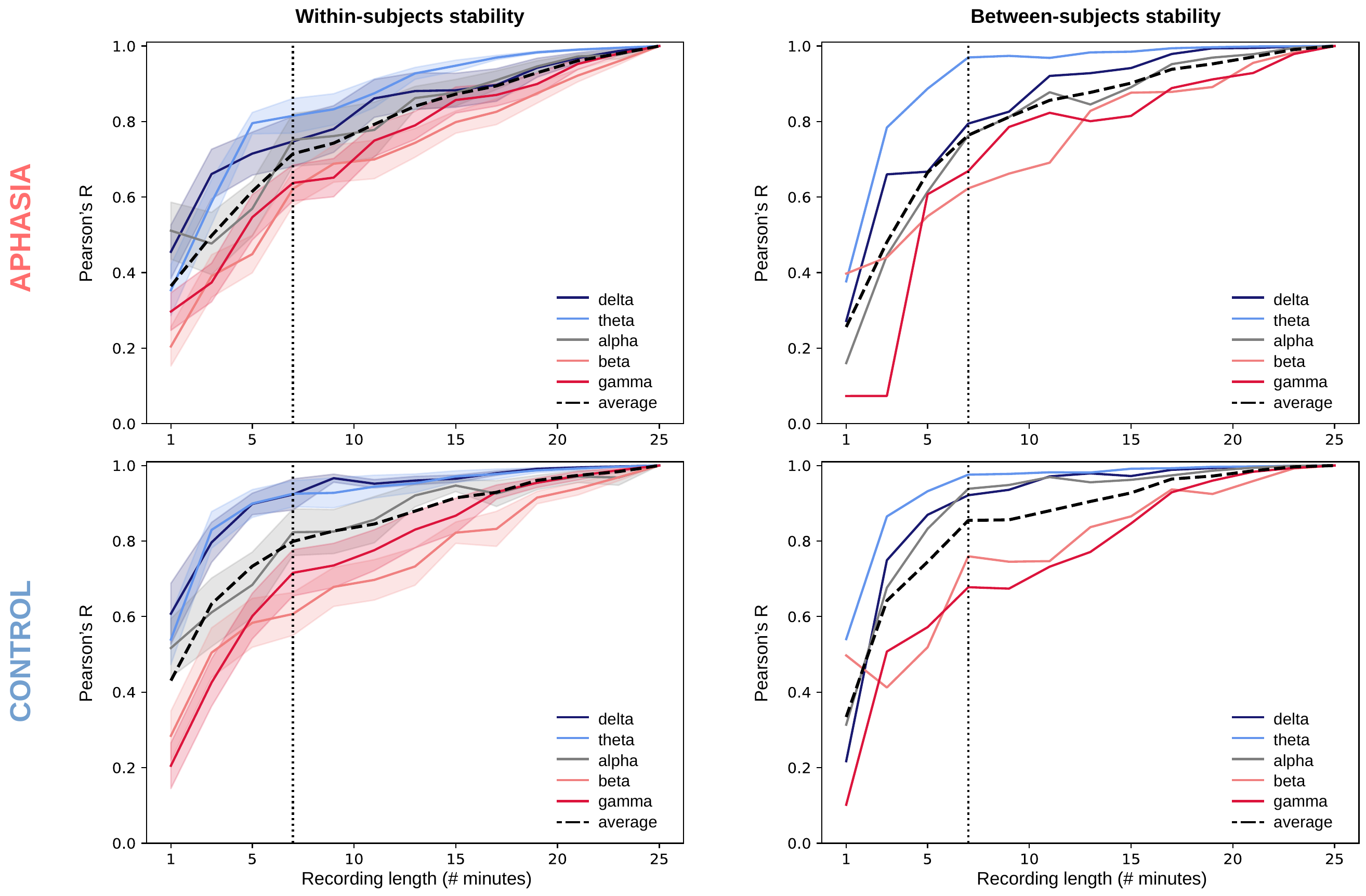}
\caption[Sigs]{\textbf{Stability measures grouped}. Within- and between-subjects stability analysis performed for each group separately.  Black dotted line indicates the average across frequencies. Shaded areas indicate the standard error of the correlations. The knee point of all panels is indicated with a vertical dotted line (based on the average across frequencies)} 
\label{fig:stab_groups}
\end{figure}

\newpage
\subsection*{Demographic information}

\setlength{\tabcolsep}{1.7pt}
\renewcommand{\arraystretch}{0.75}

\begin{table}[h]
\centering
\renewcommand{\tablename}{Supplementary Table}
\caption{Demographics and lesion information of the aphasia group}
\begin{tabular}{lllllllllll}
\hline
ID(n=27) & age & sex & \begin{tabular}[c]{@{}l@{}}Time since\\stroke\\(months)\end{tabular} & \begin{tabular}[c]{@{}l@{}}Stroke\\ type\end{tabular} & \begin{tabular}[c]{@{}l@{}}Blood\\ vessel\end{tabular} & \begin{tabular}[c]{@{}l@{}}Lesioned\\hemi-\\sphere\end{tabular} & SLT & \begin{tabular}[c]{@{}l@{}}NBT score\\ (max=276) \end{tabular} & \begin{tabular}[c]{@{}l@{}}ScreeLing \\ score (max=72) \end{tabular}\\ \hline
sub-006 & 86 & m & 22.8 & ischemia & VA & bilateral & yes & 263 & 70\\ 
sub-008 & 71 & f & 21.1 & ischemia & MCA & left & no & 273 & 69\\ 
sub-009 & 67 & m & 22.6 & ischemia & MCA & left & no & 262 & 65.5\\ 
sub-010 & 74 & m & 20.6 & ischemia & PCA & left & no & $\diamond$ & 69 \\ 
sub-014 & 75 & m & 18.2 & ischemia & MCA & left & yes & 185 & 58\\ 
sub-016 & 68 & m & 18.0 & ischemia & MCA & bilateral & yes & 265 & 69\\
sub-017 & 88 & m & 8.2 & ischemia & \begin{tabular}[c]{@{}l@{}}MCA/\\PCA\end{tabular} & bilateral &  no & 245 & 56\\ 
sub-018 & 61 & f & 29.4 & ischemia & MCA & left &  yes & 263 & 72\\ 
sub-019 & 72 & m & 8.6 & ischemia & PCA & left & $\diamond$ & 211 & 51\\
sub-020 & 42 & m & 18.7 & ischemia & MCA & left & yes & 273 & 71\\
sub-021 & 81 & m & 8.6 & hemorrhage & $\diamond$ & left &  no & 252 & 70 \\ 
sub-022 & 78 & f & 8.2 & hemorrhage & $\diamond$ & left & yes & 207 & 56\\ 
sub-023 & 90 & m & 22.8 & ischemia & LLA & bilateral &  no & 265 & 61\\ 
sub-024 & 69 & m & 6 & ischemia & MCA & left & yes &  260  & 68.5\\
sub-025 & 69 & m & 11.5 & ischemia & MCA & left & yes & 197  & 58.5\\ 
sub-026 & 71 & m & 31 & ischemia & MCA & left &  yes & 250  & 69\\
sub-027 & 76 & m & 126.2 & ischemia & MCA & left &  yes & 254  & 63\\ 
sub-028 & 80 & m & 25.2 & ischemia & $\diamond$ & left & no & 195  & 54\\
sub-029 & 75 & m & 22.6 & ischemia & MCA & left &  yes & 193  & 55\\
sub-030 & 49 & m & 13.1 & hemorrhage & $\diamond$ & left &  yes & 217  & 57\\ 
sub-031 & 79 & m & 12.9 & ischemia & MCA & left &  yes & 263  & 63\\
sub-032 & 76 & m & 94.8 & ischemia & MCA & left &  yes & 3  & 28\\
sub-034 & 79 & m & 13.4 & $\diamond$ & $\diamond$ & $\diamond$ &  yes & 244  & 69\\
sub-035 & 64 & f & 31.5 & ischemia & \begin{tabular}[c]{@{}l@{}}MCA/\\ACA\end{tabular} & left &  yes & 276  & 71\\ 
sub-038 & 60 & f & 368.6 & $\diamond$ & $\diamond$ & $\diamond$ &  no & 251  & 65\\ 
sub-049 & 85 & f & 8.3 & ischemia & PICA & left &  yes & 242  & 65.5\\ 
sub-050 & 81 & f & 8.7 & ischemia & MCA & left &  yes & 134  & 53\\ \hline
Total & \multicolumn{1}{c}{\begin{tabular}[c]{@{}l@{}}73\\ $\pm$ 11 \end{tabular}} & \multicolumn{1}{c}{\begin{tabular}[c]{@{}l@{}}20m\\7f\end{tabular}} & \multicolumn{1}{c}{\begin{tabular}[c]{@{}l@{}}37\\ $\pm$ 71 \end{tabular}} & \multicolumn{1}{c}{\begin{tabular}[c]{@{}l@{}}23 ischemia \\3 hemorrhage\end{tabular}} & \multicolumn{1}{c}{\begin{tabular}[c]{@{}l@{}}17 MCA\\3 PCA\\1 ACA\\1 VA\\1 LLA\\1 PICA\end{tabular}} & \multicolumn{1}{c}{\begin{tabular}[c]{@{}l@{}}22 left\\4 bilateral\end{tabular}} & \multicolumn{1}{c}{\begin{tabular}[c]{@{}l@{}}21 yes\\8 no\end{tabular}} & \multicolumn{1}{c}{\begin{tabular}[c]{@{}l@{}}228.58\\ $\pm$ 57.75\end{tabular}} & \multicolumn{1}{c}{\begin{tabular}[c]{@{}l@{}} 62.11\\ $\pm$ 9.43\end{tabular}}\\ \hline
\end{tabular}
\\
\begin{flushleft}
\footnotesize{SLT=speech-language therapy; NBT=Dutch naming test (Nederlandse Benoem Test); VA=vertebral artery; MCA=middle cerebral artery; PCA=posterior cerebral artery; ACA=anterior cerebral artery; LLA=lateral lenticulostriate arteries; PICA=posterior inferior cerebellar artery; $\diamond$=data not available}
\end{flushleft}
\label{table:participants}
\end{table}

\newpage
\subsection*{Correlation matrix frequency bands}
\begin{table}[h]
\renewcommand{\tablename}{Supplementary Table}
\caption{Correlation matrix neural envelope tracking }
\centering
\resizebox{\textwidth}{!}
{\begin{tabular}{ccccccccccccc}
\toprule
\multicolumn{1}{c}{} & \multicolumn{6}{c}{\textbf{IWA}} & \multicolumn{6}{c}{\textbf{Controls}}\\
\cmidrule(rl){2-7} \cmidrule(rl){8-13}
\textbf & {broad} & {delta} & {theta} & {alpha} & {beta} & {gamma} & {broad} & {delta} & {theta} & {alpha} & {beta} & {gamma}\\
\midrule
\\ broad & 1 & & & & & & 1 & & & & &\\
\\
delta & 0.79 & 1 & & & & & 0.92 & 1 & & & &\\
\\
theta & 0.15 & 0.01 & 1 & & & & 0.09 & 0.02 & 1& & &\\
\\
alpha & 0.15 & 0.07 & 0.71 & 1 & & & 0.27 & 0.15 & 0.52 & 1 & &\\
\\
beta & 0.15 & 0.20 & 0.62 & 0.61 & 1 & & 0.48 & 0.57 & 0.32 & 0.44 & 1 &\\
\\
gamma & 0.19 & 0.22 & -0.01 & 0.01 & 0.23 & 1 & 0.12 & 0.18 & -0.18 & -0.22 & 0.26 & 1\\
\\
\bottomrule
\end{tabular}}
\\
\begin{flushleft}
\footnotesize{Exploratory analysis investigating the collinearity between frequency bands. The table displays the Pearson correlations for the mean MI (integration window 0-400 ms).}
\end{flushleft}
\label{table:corr}
\end{table}
\newpage

\bibliographystyle{apalike}
\bibliography{bibliography.bib}
\end{document}